\documentclass[aps,prd,preprint,eqsecnum]{revtex4}

\usepackage{amsmath,amssymb,amsfonts,verbatim}
\usepackage[dvips]{graphicx}

\def\beqn{\begin{eqnarray}}
\def\eeqn{\end{eqnarray}}
\def\nn{\nonumber}
\def\spa#1.#2{\langle#1#2\rangle}
\def\spb#1.#2{[#1#2]}

\def\spRa#1.#2{\langle#1#2]}

\def\ang#1{|#1\rangle}
\def\bra#1{|#1]}
\def\bang#1{\left\langle#1\right|}
\def\bbra#1{[#1|}
\newcommand\ab[3]{\langle#1|#2|#3]}

\def\Pslash{\hbox{$\slash \hskip -.24cm P$}}

\def\cit#1{Ref.~\cite{#1}}
\def\rf#1{(\ref{#1})}
\def\cal#1{\mathcal{#1}}

\begin{document}
\title{On-shell recursion relations for gravity}
\author{Anthony Hall}
\affiliation{{} Department of Physics and Astronomy, UCLA\\
\hbox{Los Angeles, CA 90095--1547, USA}
\\{\tt anthall@physics.ucla.edu}
}

\date{March 3, 2008}

\begin{abstract}
We extend the argument presented by Benincasa, Boucher-Veronneau, and
Cachazo to show that graviton tree amplitudes are well behaved under
large complex deformations of the momenta of a pair of like-helicity
gravitons.  This shows that BCFW recursion relations for gravity
amplitudes can be constructed using such shifts, providing an
alternative proof to the recent one by Arkani-Hamed and Kaplan.  By
using auxiliary recursion relations the cancellations which are hidden
when using covariant Feynman diagrams become manifest.
\end{abstract}

\maketitle

\section{Introduction}
The standard Feynman diagram representation hides many
remarkable properties of on-shell graviton amplitudes.  For example,
the Kawai-Lewellen-Tye relations from string theory indicate that
tree-level gravity amplitudes are linear combinations of squared
gauge-theory amplitudes~\cite{KLT}.  Thus the three- and four-point
vertices of gauge theory generate the tree-level amplitudes derived
from the Einstein-Hilbert action, with its infinite series of
interactions.  More recently Witten pointed out that scattering
amplitudes in gauge and gravity theories have intriguing simplicity
in twistor space~\cite{Witten}. 
On-shell recursion relations~\cite{BCFrec, BCFWproof, BGKS} 
which express an amplitude as products of physical on-shell
amplitudes with fewer legs also exist in gravity theories~
\cite{Bedford, Cachazo, mhvgravity, taming}, although naive power
counting based on Feynman diagrams indicates otherwise.

The recursion relations of Britto, Cachazo, Feng, and Witten (BCFW)
produce very compact expressions for amplitudes and are obtained by
shifting a pair of the external legs to complex momenta~
\cite{BCFWproof}.  The existence of BCFW recursion for any field
theory hinges upon the deformed amplitudes vanishing when the shifted
legs are taken to infinite momentum in special complex directions, and
so, with the KLT relations in mind, it is natural to imagine that
gravity may share these high-energy properties and satisfy on-shell
recursion relations.
The known examples of on-shell recursion relations for graviton tree
amplitudes~\cite{Bedford, Cachazo, mhvgravity, taming} shows that the
high-energy behavior of gravity in special complex directions is
better behaved than one expects from the power counting of momenta in
individual Feynman diagrams.  It is surprising that, for some complex
deformations, gravity amplitudes are actually better behaved in these
high-energy limits than gauge theory.  Very interestingly,
the existence of an on-shell
recursive framework for the calculation of graviton tree amplitudes
implies that all of perturbative gravity follows from the on-shell
three vertex with complex momenta.

The surprising properties of graviton tree amplitudes have
repercussions at loop level.  Remarkable results for the maximally
supersymmetric $\cal N =8$ supergravity theory have inspired the
``no-triangle hypothesis" which claims that the bubble and triangle
graphs vanish along with additional rational terms, leaving only
scalar box integrals contributing to one-loop $\cal N =8$ supergravity
amplitudes~\cite{OneloopMHVGravity, NoTriangle, SixPointIR,
NoTriangleSixPt}.  The maximally helicity-violating class of
amplitudes~\cite{OneloopMHVGravity} provided the initial evidence, and
recent developments have bolstered the hypothesis for $\cal N =8$
supergravity.  Explicit calculations have verified the conjecture for
$\cal N =8$ supergravity at six points~\cite{NoTriangleSixPt},
seven-point amplitudes have infrared limits consistent with the
hypothesis~ \cite{SixPointIR, NoTriangleSixPt}, and beyond this
the scaling and factorization properties of amplitudes
provide evidence of the cancellations~\cite{NoTriangle,
NoTriangleSixPt,gravitycancel}.  The complete $\cal N =8$ four-point
amplitude at three loops gains improved ultraviolet behavior from
these cancellations~\cite{GravityThree}.

The observed cancellations do not appear to be due to supersymmetric constraints.
One-loop calculations in pure gravity indicate that the requisite
cancellations arise with the Einstein-Hilbert action alone, and that
the high-energy improvements originate from the rather tame scaling
behavior of graviton tree amplitudes~\cite{gravitycancel}.  Thus one
expects to understand the cancellations of ultraviolet divergences at
loop level by studying the tree-level amplitudes.

Benincasa, Boucher-Veronneau, and Cachazo (BBC)~\cite{taming} proved
that under the standard BCFW shift involving one positive and one
negative helicity graviton the amplitudes are well behaved, confirming
the existence of recursion relations for this case.  In this paper we
extend this proof to show that graviton tree amplitudes also scale
nicely under BCFW shifts of like-helicity gravitons, proving the
recursion relations for these cases as well.   An auxiliary
recursion is used to bring amplitudes into a form where the mild
high-energy behavior is manifest.  The upper bounds on the
high-energy scaling are independent of the number of external
gravitons being scattered.  

Very recently Arkani-Hamed and Kaplan~\cit{nima} gave an elegant
argument applicable to general $D$-dimensional field theories, using
the Lagrangian in a particular gauge \cite{Chalmers,Vaman}, to find
the high-energy scaling of gravity amplitudes under BCFW shifts of any
helicity.  The improved high energy behavior, relative to
power-counting expectations, is attributed to an enhanced spin
symmetry at infinite momentum.  Our paper gives an alternative $D=4$
proof that graviton amplitudes satisfy the scaling needed for BCFW
recursion relations under the like-helicity BCFW shifts not handled in
Refs.\cite{taming,gravitycancel}.

We begin by reviewing the BCFW recursion relations.  The BBC proof of the 
recursion formulas for pure gravity under the complex deformations of two 
opposite helicity gravitons are examined, particularly the use of specially 
designed auxiliary recursion formulas.  We extend this argument to the 
like-helicity BCFW case and find that graviton amplitudes vanish when 
the shift parameter is taken to infinity.

\section{Review of On-Shell Recursion}
We consider first the BCFW recursion relations for tree-level gravity 
amplitudes which are obtained by shifting a pair of spinors from the external 
gravitons $i$ and $j$, 
\beqn
	\ang{i(z)} = \ang{i} + z \ang{j},\quad 
	\bra{j(z)} = \bra{j} - z \bra{i},
\label{BCFWShift}
\eeqn
where $\ang{i}$ and $\bra{j}$ are Weyl spinors of positive and negative
helicity.  The momenta of these two gravitons are still null but
shifted to complex values,
\beqn \label{BCFWShiftp}
	p_i(z)  = \ang{i} \bbra{i}+z \ang{j} \bbra{i}, \quad
	p_j(z) =  \ang{j} \bbra{j}-z \ang{j} \bbra{i},
\eeqn
while overall momentum remains conserved,
\beqn
	p_i(z) + p_j(z) = p_i+p_j.
\eeqn
We use the shorthand ``$\spRa i.j$" to denote this choice of shifted spinors.
  
The polarization tensors for gravitons of positive and 
negative helicity, 
written in terms of arbitrary reference spinors $\ang{\mu}$ and $\bra{\mu}$,
are 
\beqn	\label{poltensors}
	{\epsilon_p^{++}}_{a\dot{a},b\dot{b}} &=& 
	\frac{(\ang{\mu}_a \bbra{p}_{\dot{a}})(\ang{\mu}_b \bbra{p}_{\dot{b}})}
	{ \spa\mu.p^2}, 
	\\ \nn
	{\epsilon_p^{--}}_{a\dot{a},b\dot{b}} &=&  
	\frac{(\ang{p}_a \bbra{\mu}_{\dot{a}}) (\ang{p}_b \bbra{\mu}_{\dot{b}})}
	{ \spb\mu.p^2}.
\eeqn
If leg $p$ is deformed under the BCFW shift, then its polarization tensor also 
gains $z$ dependence.  If, for example, leg $p$ is $i^+$ or $j^-$ we have 
$\epsilon_p(z) \sim \frac{1}{z^2}$.

After applying the BCFW shift, a gravity amplitude $M$ remains on shell but is 
now a meromorphic function $M(z)$ with simple poles in $z$ whenever a 
$z$-dependent propagator in its Feynman diagrams is on shell.  At tree level 
a propagator partitions a Feynman diagram into two trees attached at the 
internal line, so the internal line is $z$-dependent for Feynman diagrams 
which partition legs $i$ and $j$ into opposite trees.  Writing $P(z)$ for the 
sum of the external particles' momenta at one tree which flows through the 
internal line, the condition for the propagator to yield a pole is 
\beqn
	0 = P(z)^2 = P^2 + z \ab{j}{P}{i},
\eeqn
where $ \ab{j}{P}{i} \equiv \langle j^-| \Pslash |i^- \rangle$.
Provided  that $M(z)$ vanishes as $z \rightarrow \infty$, we have 
$\oint \! M(z)/z = 0$.  
Applying Cauchy's theorem to $\oint \! M(z)/z$
yields a residue for $z=0$, the desired unshifted amplitude, and residues 
for the values of $z$ for which $P(z)^2 = 0$.
The universal factorization 
of amplitudes at these poles yields the BCFW formula \cite{BCFWproof},
\beqn
	M_n\left(p_1,\ldots,p_n\right) = 
	\sum_{\cal{I},\cal{J}} \sum_{h=\pm} 
	M_{\cal{I}}\left(p_\cal{I}(z),\ldots,-P_\cal{I}^h(z)\right)
	\times \frac{1}{P_\cal{I}^2} \times 
	M_{\cal{J}}\left(P_\cal{J}^{-h}(z),\ldots,p_\cal{J}(z)\right).
   \hskip .5 cm 
\eeqn

Each term appearing in the recursion formula is one of the residues of
$M(z)/z$.
We use the index $\cal{J}$ to denote the external gravitons attached to the 
same subamplitude as the BCFW-shifted leg $j$, so that $j\in\cal{J}$, and 
$\cal{I}$ represents all the remaining external legs.
  The momentum $P_\cal{I}(z)$ is the shifted total momentum
leaving from $M_\cal{I}$'s external legs, and the recursion
formula sums over the helicity of the internal state with momentum
$P_\cal{I}$.  The subamplitudes $M_{\cal{I},\cal{J}}$ are on shell, 
evaluated with the shifted spinors at the value of $z=z_\cal{I}$ for which
$P_\cal{I}(z)^2 = 0$.

The internal on-shell momentum can be written explicitly as a two-spinor.  
From the identities 
\beqn
	P(z)\bra{i}  = P \bra{i},  \quad 
	\bang{j} P(z) = \bang{j} P,
\eeqn
where the slash on the $P$ is implicit,
the shifted spinors of an internal particle can be written as
\beqn
	P(z) &=& \ang{P(z)}\bbra{P(z)} \\ \nn
	&=& \frac{1}{\ab{j}{P}{i}} (P\bra{i}) (\bang{j} P).
\eeqn

The potential obstruction to constructing BCFW recursion relations for an arbitrary 
theory is that the amplitudes must vanish for large $z$.  In general this 
condition is not straightforward to verify.  For example, in gauge theory and 
gravity the Lagrangians provide vertices which scale like $z$ and $z^2$ at 
high energy, respectively.  To tame this high-energy behavior 
of graviton scattering amplitudes, 
and in turn prove that the BCFW shift ``$\spRa i^+.{j^-}$" indeed generates 
on-shell recursion, BBC used an auxiliary recursion \cite{taming}.

The auxiliary recursion relations follow from a shift of all the 
positive-helicity gravitons~$k^+$ and a single negative helicity leg $j$, 
\beqn \label{shift1}
\Big\{
	\ang{k(w)} &=& \ang{k} + w \ang{j}\Big\} 
	\quad \forall~ k\in {k^+}, \\ \nn
	\bra{j(w)} &=& \bra{j} -w \sum_{k\in {k^+}} \bra{k}.
\eeqn
The momentum of an internal propagator after applying the auxiliary shift is 
\beqn
\label{Phatw}
	P_\cal{I}(w) = P_\cal{I} + w \ang{j} \sum_{k\in \cal{I^+}} \bbra{k},
\eeqn
so that the values of $w$ which yield an on-shell internal propagator are 
\beqn
\label{onshellw}
	w_\cal{I} = - \frac{P_\cal{I}^2}
		{\sum_{k\in \cal I^+}\ab{j}{P_\cal{I}}{k}}.
\eeqn
We use the conventions of \cit{taming} where 
$P_\cal{I}=\sum_{k\in \cal{I}} p_k$ and define $\cal{I^+}$ to be the set of 
positive-helicity external gravitons in $M_\cal{I}$.

A non-vanishing contribution to the recursion can only be 
achieved when the on-shell internal propagator 
partitions the Feynman diagram such that at least one of the positive helicity 
gravitons is on the opposite side of the propagator from the others.
Thus the auxiliary recursion generates diagrams which partition 
the external legs into $\cal{I}$ and $\cal{J}$ labeled 
such that $j\in \cal{J}$ in correlation with the BCFW shift, and at least one 
positive-helicity graviton lies in $\cal{I}$ and another in $\cal{J}$.  

As in the BCFW case, the internal on-shell momentum can be written explicitly 
as a two-spinor.  
From the identities 
\beqn
	P_{\cal{I}}(w)\bra{\alpha} &=& P_{\cal{I}}\bra{\alpha} 
		\mbox{ for }
	\bra{\alpha} = \sum_{k\in \cal{I^+}} \bra{k}, \\ \nn
	\bang{j}P_{\cal{I}}(w) &=& \bang{j}P_\cal{I},
\eeqn
the shifted spinors of an internal particle can be written as
\beqn \label{phat}
	P_\cal{I}(w) &=& \ang{P_\cal{I}(w)} \bbra{P_\cal{I}(w)} \\ \nn
	&=& \frac{1}{\ab{j}{P_\cal{I}}{\alpha}} 
	(P_\cal{I} \bra{\alpha}) (\bang{j} P_\cal{I}).
\eeqn

To have the proper behavior to yield on-shell recursion for gravity amplitudes 
the auxiliary shift must make an $n$-point amplitude
$M_n(w)$ vanish as $w$ is taken to infinity.  The most divergent Feynman 
diagrams contributing to $M_n(w)$ contain only cubic vertices, $n-2$ of them, 
each contributing a factor of $w^2$.  The $p$ positive-helicity gravitons and 
the negative helicity leg $j$ each yield a factor of $1/w^2$ from their 
polarization tensors for a total of $1/w^{2p+2}$.  Also the $w$-dependence in 
each of the $n-3$ propagators 
cannot cancel in general \cite{taming}, so $1/w^{n-3}$ is gained from propagators.  
Overall the 
most divergent contributions to $M_n(w)$ scale at large $w$ as $1/w^{p-m+3}$, 
where we write $p$ and $m$ for the number of external gravitons of positive- 
and negative-helicity, respectively. 
Thus the above auxiliary shift leads to on-shell recursion relations for gravity 
amplitudes with $p-m\geq -2$.

Likewise, gravity amplitudes with $m-p\geq -2$ vanish at large $w$  
under a shift of all the negative-helicity gravitons $k^-$ 
and a single positive helicity leg $i$,
\beqn \label{shift1mreg}
\Big\{\bra{k(w)} &=& \bra{k} - w \bra{i}\Big\} 
	\quad \forall~ k\in {k^-}, \\ \nn
	\ang{i(w)} &=& \ang{i} +w \sum_{k\in {k^-}} \ang{k}.
\eeqn
The set of spinor deformations introduced by BBC give recursion
relations for all graviton scattering amplitudes, and thus allow
gravity amplitudes to be recursively calculated starting from on-shell, 
complex three vertices
alone.  However, these recursion relations are overly complicated
compared to the relations obtained using the standard BCFW shift
(\ref{BCFWShift}).  They are, however, useful for proving 
that the standard shifts (\ref{BCFWShift}) lead to good recursion relations. 

\section{Review of the BBC proof for ``$\spRa i^+.{j^-}$" shifts}
In order to obtain recursion relations for amplitudes under the BCFW
shift we seek a representation of amplitudes where
the $z$-dependence of polarization tensors will help tame divergent
contributions from vertices.  We first review the BBC proof of the
large-$z$ behavior where leg $i$ is a positive-helicity graviton and
leg $j$ is negative.  As described in their paper, auxiliary
recursions based on the shifts described in the previous section are
very helpful for showing that the amplitude vanishes for large $z$
after a BCFW shift, eqn.~(\ref{BCFWShift}).  By applying the auxiliary shifts
of eqns.~(\ref{shift1}) and (\ref{shift1mreg})
the associated recursion relations, valid for all graviton amplitudes, 
give $w$-dependence to each positive- or negative-helicity graviton, 
respectively.  For each
auxiliary recursion diagram, which partitions the amplitude into a
pair of tree amplitudes connected by an internal graviton, Cauchy's
theorem freezes the value $w=w_\cal{I}$ so that the internal graviton
is on shell and thus yields a pole from its propagator.

\begin{figure}
\centering
\includegraphics{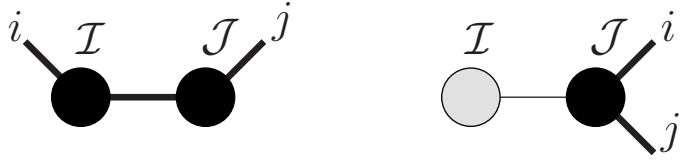}
\caption{After applying a BCFW shift to the auxiliary diagrams, legs $i$ 
and $j$ lie on either opposite sides of the internal propagator or both on 
the same tree amplitude.  $M_\cal{J}$ is labeled as the amplitude with the 
BCFW-shifted leg $j$, $M_\cal{I}$ has the remaining external gravitons.  The 
other external gravitons besides $i$ and $j$ are suppressed. The $z$-dependent 
gravitons and on-shell amplitudes are indicated in black.
\label{aux0}}
\end{figure}

Now consider applying the BCFW shift ``$\spRa i^+.{j^-}$" to the auxiliary 
diagrams.  Gravitons $i$ and $j$ either lie on opposite sides of the internal 
graviton, or they are both part of the tree amplitude $M_\cal{J}$.  
The two possibilities are illustrated in Fig.~\ref{aux0}.

If an auxiliary diagrams has $i$ and $j$ on opposite sides of the partition, 
then $P_\cal{I}$ is shifted to $P_\cal{I}(z)=P_\cal{I}+z\ang{j}\bbra{i}$ and 
the values of $w$ at the auxiliary poles for the deformation 
eqn.~(\ref{shift1}) are shifted to be just linear in $z$,
\beqn
\label{wz}
	w_\cal{I}(z) &=& - \frac{P_\cal{I}^2+z \ab{j}{P_\cal{I}}{i}}
 	{\sum_{k\in \cal I^+}\ab{j}{P_\cal{I}}{k}} \\ \nn
	&=& w_\cal{I} - 
	z \frac{\ab{j}{P_\cal{I}}{i}}{\sum_{k\in \cal I^+}\ab{j}{P_\cal{I}}{k}}.
\eeqn
The internal momentum can be written as in eq.~\rf{phat} so that after the 
BCFW shift with $z$ we have
\beqn
	P_\cal{I}(w(z),z) = 
	&=& \frac{1}{\ab{j}{P_\cal{I}}{\alpha}} 
	\Big((P_\cal{I}+z\ang{j}\bbra{i}) \bra{\alpha}\Big) (\bang{j} P_\cal{I}).
\eeqn
The $z$-dependence of the internal momentum's spinors is confined to 
$\ang{P_\cal{I}(w(z),z)}$.

Provided that leg $i$ is not the only positive-helicity graviton on
$M_\cal{I}$, all the polarization tensors deformed by the auxiliary shift
help to reduce the degree of divergence under the BCFW shift.  The
most divergent Feynman diagrams for $M_\cal{I}(w(z),z)$ contain only
three vertices which each contribute $z^2$.  Let $m_\cal{I}$ and $p_\cal{I}$ 
denote the number of negative- and positive-helicity external gravitons for 
$M_\cal{I}$.  Including the internal
graviton there are $m_\cal{I}+p_\cal{I}+1$ particles in $M_\cal{I}$
and thus would have $m_\cal{I}+p_\cal{I}-1$ three vertices giving a
total of $z^{2(m_\cal{I}+p_\cal{I}-1)}$ from vertices.  The
$p_\cal{I}$ positive helicity gravitons and the $\ang{P(z)}$ from the
internal particle contribute $1/z^{2(p_\cal{I}+h)}$ from polarization
tensors.  With all $m_\cal{I}+p_\cal{I}-2$ propagators depending on
$z$ as shown in \cit{taming} the propagators yield
$1/z^{m_\cal{I}+p_\cal{I}-2}$.  Overall we have
\beqn
	M_\cal{I}(w(z),z) \sim 1/z^{p_\cal{I}-m_\cal{I}+2h}.
\eeqn

The same counting for $M_\cal{J}(w(z),z)$ gives an extra factor of $1/z^2$ from 
the polarization tensor of leg $j$ and $1/z^{-2h}$ from the internal particle's 
polarization.  Overall we have 
\beqn
	M_\cal{J}(w(z),z) \sim 1/z^{p_\cal{J}-m_\cal{J}-2h+2}.
\eeqn

Combining the contributions of $M_\cal{I}(w(z),z)$, $M_\cal{J}(w(z),z)$, and the 
internal propagator $1/P_\cal{I}(z)^2$ to the auxiliary diagrams with 
$i$ and $j$ on opposite trees, the leading $z$ behavior for this class of 
auxiliary diagrams is $1/z^{p-m+3}$ and similarly for auxiliary diagrams from 
the shifts of eqn.~(\ref{shift1mreg}).  The $z$ dependence of polarization 
tensors produced through the auxiliary recursion has thus tamed the divergent 
contributions from three vertices.

In case leg $i$ is the only external graviton with positive helicity on its 
partition, $\cal{I^+}=\{i\}$, then $w(z)=w-z$ which gives 
\beqn
	\ang{i(w(z),z)}&=&\ang{i(z)}+w(z)\ang{j} = \ang{i}+z\ang{j}+(w-z)\ang{j}
	 = \ang{i(w)}, \\ \nn
	P_\cal{I}(w(z),z) &=& 
	\frac{1}{\ab{j}{P_\cal{I}}{i}} 
	(P_\cal{I} \bra{i}) (\bang{j} P_\cal{I}),
\eeqn
so that $M_\cal{I}$ is actually independent of $z$.  The tree 
$M_\cal{J}$ has the deformed spinors
\beqn
	\bra{j(w(z),z)} &=& \bra{j(z)}-w(z)\sum_{k\in {k^+}} \bra{k}=
	 \bra{j}-z\bra{i}-w\sum_{k\in {k^+}} \bra{k}+z\sum_{k\in {k^+}} \bra{k} 
	\\ \nn &=& 
	 \bra{j(w)}+z\sum_{k\in {\cal{J}^+}} \bra{k}, \\ \nn
	\Big\{ \ang{k(w(z))} &=& \ang{k}+w(z)\ang{j}=\ang{k(w)}-z\ang{j} \Big\}
	 \quad \forall~ k\in\cal{J}^+.
\eeqn
Provided that the internal graviton leaving tree $\cal{J}$ has negative 
helicity, this is precisely the auxiliary shift of all the positive helicity 
gravitons again, where the parameter $w$ in eqn.~(\ref{shift1}) 
has been replaced by $-z$.  The tree $\cal{J}$ contains a 
sufficient number of positive-helicity gravitons to ensure that 
$1/P_\cal{I}(z)^2 \times M_\cal{J}(z)$ vanishes at large $z$. 
If instead the internal graviton leaves tree $\cal{I}$ with negative helicity, 
then $M_\cal{I}$ is either a three vertex vanishing under the BCFW shift 
or trivially vanishes with too many negative-helicity legs.  
Similar arguments apply to the BCFW-shifted diagrams obtained from the 
auxiliary shifts of eqn.~(\ref{shift1mreg}).

Auxiliary diagrams for which the BCFW-shifted legs $i$ and $j$ 
both lie on the same tree, $i\in\cal{J}$, give $w(z)=w$ and from 
eqn.~(\ref{shift1}) we have 
\beqn
	\ang{i(w,z)} &=& \ang{i(z)}+w\ang{j} = \ang{i(w)}+z\ang{j}, \\ \nn
	\bra{j(w,z)} &=& \bra{j(z)}-w\sum_{k\in {k^+}} \bra{k} = \bra{j(w)}-z\bra{i}.
\eeqn
Thus in this class of auxiliary diagrams, the net effect of both shifts is a 
BCFW deformation of the amplitude $M_\cal{J}(w)$.  Similar statements are true 
for the auxiliary diagrams obtained from eqn.~(\ref{shift1mreg}).  By 
repeated application of 
the auxiliary recursion any graviton tree amplitude can be reduced to products 
of on-shell three vertices.  In the process of reducing an amplitude into 
products of 
subamplitudes via the auxiliary recursion, one encounters partitions of 
the legs $i$ and $j$ which fall into one of the classes outlined above.  
The special class in which $i$ and $j$ both lie on $M_\cal{J}$ recursively 
yields a BCFW-shifted on-shell three vertex, 
\beqn
	M(P^\pm,i^+(z),j^-(z))	\sim 1/z^2.
\eeqn
Thus graviton tree amplitudes under the ``$\spRa i^+.{j^-}$" shift vanish 
at large $z$ and are dominated by the three vertices $M(P^\pm,i^+(z),j^-(z))$ 
scaling like $1/z^2$.

\section{Gravity amplitudes under ``$\spRa i^-.{j^-}$" shifts}
\label{mm}
Now we use the BBC auxiliary recursion to determine the scaling of graviton 
amplitudes under the like-helicity BCFW shifts, first considering the 
``$\spRa i^-.{j^-}$" case.
Using the auxiliary recursion shown to vanish as $1/w^{p-m+3}$ in
 \cit{taming}, the shifted external spinors are 
given by eqn.~(\ref{shift1}).  This is the same as the auxiliary 
shift introduced in \cit{taming}, but leg $i$ is no longer one of the positive 
helicity legs.
The large-$w$ behavior is the same as 
\cit{taming} since we still have each of the $p$ positive-helicity polarization 
tensors contributing $1/w^2$ and the negative helicity leg $j$ gives another 
$1/w^2$ from $\bra{j(w)}$ for a total of $1/w^{2p+2}$ from polarization tensors.
Thus we have a valid auxiliary recursion for $p-m \geq -2$.
The shifted momentum of an internal propagator is given by eqn.~(\ref{Phatw})
so that the values of $w$ in eqn.~(\ref{onshellw}) yield an on-shell internal 
propagator.

For graviton amplitudes with $p-m\leq-2$ we use the following auxiliary shift. 
The negative-helicity spinors of all the negative helicity legs $k^-$ 
are shifted, except for leg $i^-$,
\beqn \label{shift1m}
\Big\{
	\bra{k(w)} &=& \bra{k} - w \bra{i}
	\Big\} \quad \forall~ k\in k^-\setminus i, \\ \nn
	\ang{i(w)} &=& \ang{i} + w \sum_{k\in {k^-\setminus i}} \ang{k}.
\eeqn
Graviton amplitudes scale as $1/w^{m-p-1}$ at large $w$ under this auxiliary 
shift because the negative-helicity leg $i$ contributes a factor 
of $w^2$ from its polarization tensor.  Thus for amplitudes with $p-m\leq-2$ 
the shift in 
eqn.~(\ref{shift1m}) gives auxiliary recursion relations, and if $p-m\geq-2$ then 
eqn.~(\ref{shift1}) is applicable.  The combination of both auxiliary shifts can 
reduce any graviton amplitude to products of on-shell three vertices.

Consider applying the BCFW shift 
of two negative-helicity legs $i$ and $j$, 
\beqn
	\ang{i(z)} = \ang{i} + z \ang{j},\quad 
	\bra{j(z)} = \bra{j} - z \bra{i},
\eeqn
to an amplitude $M(w)$.  
Auxiliary diagrams, as in Fig.~\ref{aux0}, for which $i$ and $j$ lie on opposite sides of the 
propagator have $w_\cal{I}(z)$ linear in $z$ as in eqn.~(\ref{wz}), 
and as in \cit{taming} there are no accidental cancellations which remove 
the $z$-dependence of any of the Feynman diagram propagators.  The most 
divergent contributions to $M_\cal{I}(w(z),z)$ and $M_\cal{J}(w(z),z)$ arise 
from Feynman diagrams with only three vertices that each scale as $z^2$.
In the cases with $p-m\geq-2$, auxiliary diagrams with $i$ and $j$ on opposite 
sides of the propagator gain 
a factor of $z^2$ relative to the ``$\spRa i^+.{j^-}$" case from $\ang{i(z)}$ 
in the polarization tensor of leg $i$.  This gives such 
auxiliary diagrams a $z$-dependence at large $z$ which is $1/z^{p-m+1}$, 
vanishing for $p\geq m$.  
In the cases with $p-m\leq-2$, $\ang{i(w(z),z)}$ gives the auxiliary diagrams 
with $i$ and $j$ on opposite sides of the propagator a factor of $z^4$ relative 
to the scaling of the ``$\spRa i^+.{j^-}$" case.  These auxiliary diagrams 
scale as $1/z^{m-p-1}$ for large $z$, vanishing for $m-p\geq2$.  
The special class of auxiliary diagrams for which $\cal{J}^-=\{j\}$, similar 
to the $\cal{I}^+=\{i\}$ cases for the ``$\spRa i^+.{j^-}$" argument, are 
relevant for the $m-p\geq2$ auxiliary diagrams and also vanish at large $z$.

\begin{figure}
\centering
\includegraphics{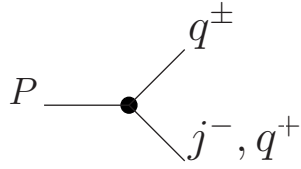}
\caption{The vertices in $M_\cal{J}(w(z),z)$, with external gravitons $j$ and 
$q$ and internal line $P$, which scale at large $z$ better than expected from 
power counting. 
\label{vertex}}
\end{figure}

Unlike the situation for ``$\spRa i^+.{j^-}$" BCFW shifts, the auxiliary 
recursion relations do not directly give vanishing 
large-$z$ behavior in the ``$\spRa i^-.{j^-}$" cases where $i$ and $j$ lie on 
opposite sides of the propagator.  Using the naive $z^2$ scaling for three 
vertices, the auxiliary diagrams scale like a constant at large $z$ for the case 
$p-m=-1$.  However, we have used the three vertex in 
de Donder gauge~\cite{sannan, dewitt} with two on-shell legs
to find that the large-$z$ scaling
for the three vertex to which two external legs attaches is better than the 
naive scaling.  We find that the large-$z$ scaling of the vertex in 
Fig.~\ref{vertex} is improved by the factor $1/z$ for auxiliary diagrams 
with $i\notin\cal{J}$ under the auxiliary shift of eqn.~(\ref{shift1}).
If a positive helicity leg $q^+$ is attached to this three vertex 
it has $z$ dependence through $w(z)$ under the shift of eqn.~(\ref{shift1}).
The momentum $q^-$ is independent of $z$, but $q^+$ is shifted to 
\beqn
	q(w_\cal{I}(z)) = q(w_\cal{I}) - z 
	\frac{\ab{j}{P_\cal{I}}{i}}{\sum_{k\in \cal I^+}\ab{j}{P_\cal{I}}{k}} 
	\ang{j} \bbra{q}.
\eeqn
The momentum $p_j$ after applying the auxiliary and BCFW shifts is 
\beqn
	p_j(w_\cal{I}(z),z) = p_j(w_\cal{I}) - z \ang{j} 
	\left( \bbra{i} - 
	\frac{\ab{j}{P_\cal{I}}{i}}{\sum_{k\in \cal I^+}\ab{j}{P_\cal{I}}{k}}
	 \sum_{k\in \cal I^+} \bbra{k}
	\right).
\eeqn

We take the polarization vectors for the shifted legs $p_j^-$ and $q^\pm$ to be 
\beqn
	\epsilon_j^- = \epsilon_q^+ = \ang{j} \bbra{q}, \quad\quad
	\epsilon_q^- = \ang{q} \bbra{j},
\eeqn
where $\epsilon_q^-$ need not shift under the deformations to remain orthogonal to $q^-$.  
The graviton polarization tensors are symmetric tensor products of these 
vectors, normalized as in eqn.~(\ref{poltensors}).
The possible vertices with two external $z$-dependent legs to consider are 
\beqn
V_{\mu \alpha,~\nu \beta,~\sigma \gamma}(p_j^-(w_\cal{I}(z),z), q^-, P(z)), \\ 
V_{\mu \alpha,~\nu \beta,~\sigma \gamma}(p_j^-(w_\cal{I}(z),z), q^+(w_\cal{I}(z)), P(z)), \\ \nn
V_{\mu \alpha,~\nu \beta,~\sigma \gamma}(q_1^+(w_\cal{I}(z)), q_2^+(w_\cal{I}(z)), P(z)), \\ \nn
V_{\mu \alpha,~\nu \beta,~\sigma \gamma}(q_1^-, q_2^+(w_\cal{I}(z)), P(z)).
\eeqn
The leading $z^2$ terms in these vertices cancel after contracting the vertex 
with polarization tensors for the external legs, 
$\epsilon_j^{--}$ and $\epsilon_q^{\pm\pm}$.

  Above we 
assumed the vertex contributed $z^2$ with factors of $1/z^2$ from the 
polarization tensors $\epsilon_j^{--}$ and $\epsilon_q^{++}$, but the calculation 
from the three vertex shows that only the subleading terms proportional to 
$z$ contribute.
This improves the estimate for the large-$z$ scaling of auxiliary diagrams 
with $i$ and $j$ on opposite sides of the propagator by a factor $1/z$, 
yielding $1/z^{p-m+2}$ behavior at large $z$ under ``$\spRa i^-.{j^-}$" for 
this type of auxiliary diagram when $p-m\geq-1$.

Auxiliary diagrams for which $i$ and $j$ lie on the same 
side of the propagator, $i \in \cal{J}$, give $w_\cal{I}(z)=w_\cal{I}$ and for 
the auxiliary shifts in eqns.~(\ref{shift1}) and (\ref{shift1m}) we have  
\beqn
	\ang{i(w,z)} = \ang{i(w)} + z \ang{j},\quad 
	\bra{j(w,z)} = \bra{j(w)} - z \bra{i},
\eeqn
As in the previous ``$\spRa i^+.{j^-}$" case, we have $z$ dependence only in 
$M_\cal{J}(w)$ as a BCFW-shifted on-shell amplitude.  To determine the 
large-$z$ scaling of this subamplitude, we apply the auxiliary recursion to 
$M_\cal{J}$ by deforming with the appropriate auxiliary shift in 
eqns.~(\ref{shift1}) and (\ref{shift1m}).

\begin{figure}
\centering
\includegraphics{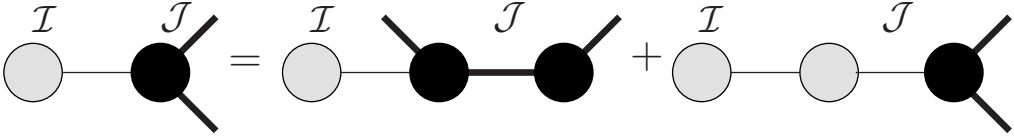}
\caption{An auxiliary diagram with $i\in\cal{J}$ on the left-hand side is 
drawn on the right-hand side after the auxiliary recursion 
has been applied to $M_\cal{J}$. 
\label{aux}}
\end{figure}

Applying the auxiliary recursion to $M_\cal{J}$ yields auxiliary subdiagrams 
belonging to the class where 
either $i$ and $j$ are on opposite sides of the internal propagator or both 
$i$ and $j$ are on the same side.  The situation after applying the 
auxiliary recursion relations to $M_\cal{J}$ is depicted in Fig.~\ref{aux}.

After 
repeated application of the auxiliary shift to the subamplitudes in the 
latter subset of diagrams, eventually we reach an on-shell three vertex,
\beqn
	M(P^+,i^-(z),j^-(z)) = 
	\left(\frac{\spa i.j^3}{\spa j.P \spa P.{i(z)} } \right)^2
	\sim 1/z^2,
\eeqn
vanishing for large $z$.
The auxiliary 
subdiagrams with $i$ and $j$ on opposite sides of the internal propagator 
have already been discussed above and vanish at large $z$.
We conclude that all graviton amplitudes satisfy  the 
``$\langle i^-,j^-]$" BCFW recursion since all the auxiliary diagrams vanish 
at large $z$ under the BCFW shift.

\section{Gravity amplitudes under ``$\spRa i^+.{j^+}$" shifts}
Because we have shown that graviton scattering amplitudes vanish 
at large $z$ after applying the ``$\spRa i^-.{j^-}$" BCFW shift, the same 
result for amplitudes under the ``$\spRa i^+.{j^+}$" shift follows by 
applying a parity reflection.  
Alternatively, auxiliary recursion relations can be used to calculate the 
large-$z$ behavior for a ``$\spRa i^+.{j^+}$" shift.  The auxiliary shifts 
used are
\beqn\label{shift2}
\Big\{
	\ang{k(w)} &=& \ang{k} + w \ang{j}
	\Big\} \quad \forall~ k\in k^+\setminus j, \\ \nn
	\bra{j(w)} &=& \bra{j} -w \sum_{k\in {k^+\setminus j}} \bra{k},
\eeqn
shifting the positive-helicity spinors of all the positive-helicity gravitons 
but $j$, and the shifts in eqn.~(\ref{shift1mreg}).
The analysis of the auxiliary diagrams after 
applying the ``$\spRa i^+.{j^+}$" BCFW shift is very similar to the above  
``$\spRa i^-.{j^-}$" case.

\section{Conclusion}
We have demonstrated that a set of auxiliary recursion relations bring
graviton amplitudes $M$ into a form where it is manifest that
$\lim_{z\rightarrow\infty} M(z) = 0$ under the BCFW shift of
like-helicity gravitons, extending the argument of Benincasa,
Boucher-Veronneau, and Cachazo for the ``$\spRa{i^+}.{j^-}"$
shift~\cite{taming}.  The vanishing of graviton amplitudes at large
$z$ agrees with the more general arguments of \cit{nima}, where the
dominant contributions under large BCFW shifts are obtained from the
Lagrangian in a special gauge.  Our work gives an alternative proof of
the vanishing large-$z$ behavior when the shifted legs are of like
helicity.  We believe these results for high-energy graviton tree
amplitudes will help shed light on the mysterious ultraviolet
cancellations observed at loop level.

\section*{Acknowledgements}
The author thanks Zvi Bern and Harald Ita for many helpful discussions.  The 
author also thanks Academic Technology Services at UCLA for computer support.

\begingroup\raggedright\endgroup

\end{document}